\documentclass{article}
\usepackage{dcolumn}
\usepackage{amsmath}
\usepackage{amssymb}
\usepackage{bm}
\usepackage{graphicx}
\usepackage{epstopdf}
\usepackage{color}

\newtheorem{prop}{Proposition}

\begin{document}

\begin{center}

{\Large{Power law dependence in a random differential equation}}

\vspace{0.5cm}

Qiuxin LV, \quad Wei MA

School of Mathematical Sciences, Tiangong University, 
 Tianjin 300387, China

Jinzhi LEI\footnote{Correspondence author: jzlei@tiangong.edu.cn}

School of Mathematical Sciences, 
Center for Applied Mathematics, 
Tiangong University,
 Tianjin 300387, China

 \end{center}

\begin{abstract}
This paper studies a random differential equation with random switch perturbations.  We explore how the maximum displacement from the equilibrium state depends on the statistical properties of time series of  {the} random switches. We show a power law dependence between the upper bound of displacement and the frequency of random perturbation switches, and the slope of power law dependence is dependent on the specific distribution of the intervals between switching times.  This result  {suggests} a quantitative connection between frequency modulation and amplitude modulation under random perturbations. 

\end{abstract}
\vspace{1.0cm}

\section{Introduction}
\noindent Brownian-like motion can  {be produced from a deterministic process (also termed as \textit{deterministic Brownian motion}), which show statistical properties similar to those of Brownian motions\cite{LeiMackeyPRE2011}. For example, deterministic Brownian motion can } arise when a particle is subject to impulsive kicks, whose dynamics are modeled by the equation of form\cite{Beck:1991vo,Chew:2002ul,Mackey:2006ev}
\begin{equation}
\label{eq:1}
\left\{
\begin{aligned}
\frac{d x}{d t} &= v\\
m \frac{d v}{d t} &= - \gamma v + f(t).
\end{aligned} 
\right.
\end{equation}  
The fluctuating force $f(t)$ consisting of a sequence of $\delta$-function-like impulses given by, for example, 
\begin{equation}
f(t) = \kappa \sum_{n=0}^\infty \xi(t) \delta(t - n \tau),
\end{equation} 
and $\xi$ is a ``highly chaotic'' deterministic variable generated by $\xi(t + \tau) = T(\xi(t))$, where $T$ is an exact map or semidynamical system, \textit{e.g.}, the tent map on $[-1, 1]$ \cite{Mackey:2006ev}. Alternatively, deterministic Brownian motion can also be generated from a continuous time description of the `random force' $f(t)$ that is dependent on the state (velocity) of a particle with a lag time $\tau$, \textit{i.e.}, 
\begin{equation}
f(t) = F(v(t - \tau)),
\end{equation}
with appropriate properties of the force function $F$ \cite{10.1007/bf00275162,10.1103/physreva.35.328,LeiMackeyPRE2011}.  In  {previous studies}\cite{LeiMackeyPRE2011}, it was shown that deterministic Brownian motion can be generated from a differential delay equation of form 
\begin{equation}
\label{eq:dde1}
\begin{aligned}
\frac{d v}{d t} &= - \gamma v + \sin(2\pi \beta v(t-1)),\\
v(t) & = \phi(t), \quad - 1 \leq t \leq 0.
\end{aligned}
\end{equation}
The parameter $\beta$ measures the frequency of the nonlinear function and is essential for the property of the generated deterministic Brownian motion. Particularly, while we introduced the upper bound of the solution $v_\beta(t; \phi)$ for the equation \eqref{eq:dde1} (here we set $\gamma = 1$ without lost of generality), 
$$
K(\beta; \phi) = \lim_{T \to \infty} \sup_{t\geq T} |v_\beta(t; \phi)|,
$$
A conjecture was proposed based on numerical simulations that 
\begin{equation}
\label{eq:hypo2}
\lim_{\beta \in I, \beta \to \infty} \beta^{1/2}K(\beta; \phi)
\end{equation}
exists, independent of the initial condition $\phi$, and is positive\cite{LeiMackeyPRE2011}. Here, $I\subset (+\frac{1}{2\pi}, +\infty)$ so that the solution $v_\beta(t; \phi)$ is chaotic when $\beta \in I$.  This conjecture indicates that, in general, $K(\beta; \phi) \sim \beta^{-1/2}$ when $\beta \to \infty$. Since  $\beta$ is associated with the frequency of changing the `random force' $f(t)=\sin(2\pi\beta v(t-1))$, we asked how general is this power law dependence between the upper bound of particle  {velocities} and the  {frequencies} of the impulsive kicks.    

 {Based on the equations of form \eqref{eq:1} and \eqref{eq:dde1}, we consider} a random equation of form
\begin{equation}
\label{eq:1}
\left\{
\begin{aligned}
\dfrac{d v}{d t} &= -\gamma v + F(t)\\
x(0) &= 0
\end{aligned}
\right.
\end{equation}
where $\gamma > 0$, $F(t)$ is a `random force' that has a form of telegraph process  and randomly switches over $S = \{-1, 1\}$. For the equation \eqref{eq:1}, given a process $F(t)$, the solution of \eqref{eq:1} is deterministic, which is written as $v(t; F)$  {hereafter}. Let $K_F$ denote the upper bound of the solution $v(t; F)$ over $t\in \mathbb{R}^+$, \textit{i.e.},
$$
K_F = \lim_{T\to \infty}\sup_{t\geq T}|v(t; F)|,
$$
then $K_F$ is dependent on the process $F$. This paper studies how the upper bound $K_F$ depends on the statistical properties of the random process $F(t)$.

The random process $F(t)$ can be defined by a random sequences $\{t_k\}_{k=0}^\infty$ that satisfies
\begin{equation}
\label{eq:tk}
0  = t_0 < t_1 < t_2 < \cdots < t_k < t_{k+1} < \cdots.
\end{equation}
Given a time series $\{t_k\}_{k=0}^\infty$, the function $F(t)$ is defined as
\begin{equation}
\label{nomdefineF}
F(t) = \left\{
\begin{array}{ll}
1, \quad& t_{2k}\leq t < t_{2k+1}\\
-1, & t_{2k+1} \leq t < t_{2(k+1)}
\end{array}
\right.\qquad (k=0,1,2,\cdots).
\end{equation}
Let
\begin{equation}
\label{eq:dk}
d_k = t_{k+1} - t_k,
\end{equation}
then $d_k > 0$, and  {the positive sequence $\{d_k\}_{k=0}^\infty$ is determined by the random sequence $\{t_k\}$. Alternatively, given a non-zero sequence $\{d_k\}_{k=0}^\infty$, we can define the corresponding time sequence $\{t_k\}$ through the iteration 
\begin{equation}
t_0 = 0,\quad t_{k+1} = t_k + d_k.
\end{equation}
Therefore, the random force $F(t)$ is defined by the positive sequence $\{d_k\}_{k=0}^\infty$. 
}

Moreover, while we assume that $d_k$ are random numbers satisfying an identical distribution with a density function $\psi(d)$, the statistical properties of $F(t)$ is determined by the density function $\psi(d)$. Hence, to explore how $K_F$ depends on the statistical properties of $F$, we only need to find how $K_F$ depends on the density function $\psi(d)$. In this study, we investigate the dependence of $K_F$ with various forms of the density function $\psi(d)$, and try to conclude a generic relation between the velocity of a Brownian-like particle and the statistical properties of the random force.

\section{Results}

\subsection{Iterative formula}

Given the equation \eqref{eq:1} with $F(t)$ defined by \eqref{nomdefineF}, the solution $v(t; F)$ is well defined.  Hereafter, we always assume $\gamma  = 1$. For the sequences $\{t_k\}$ and $\{d_k\}$ defined by \eqref{eq:tk} and \eqref{eq:dk}, let $v_k = v(t_k; F)$, it is easy to solve \eqref{eq:1} to have
\begin{equation}
\label{eq:vt}
v(t; F) = \left\{
\begin{array}{ll}
1 + (v_{2k} - 1) e^{-(t - t_{2k})},\qquad & t_{2k} \leq t < t_{2k+1},\\
-1 + (v_{2k+1} +1) e^{-(t-t_{2k+1})}, \qquad& t_{2k+1} \leq t < t_{2(k+1)}.
\end{array}
\right.
\end{equation}
Here, we note $v_0 = 0$. From \eqref{eq:vt}, it is straightforward to have $-1 \leq v(t; F) \leq 1$, and $v(t; F)$  monotonously increase when $t_{2k} \leq t  < t_{2k+1}$, and monotonously decrease when $t_{2k+1} \leq t < t_{2(k+1)}$. Hence, to obtain the upper bound of the solution $v(t; F)$, we only need to examine the values at  {the time points} $t = t_k$. 

From \eqref{eq:vt}, we have the iterative formula
  \begin{equation}
  \label{exp-x}
  \left\{
  \begin{array}{rcl}
  v_{2k+1}&=&1 + (v_{2k} -1) {e^{-d_{2k}}},\\
  v_{2k + 2}  &= &-1 + (v_{2k+1} +1) {e^{-d_{2k+1}}},
  \end{array}
  \right.\quad k\geq 0
  \end{equation}
Hence, 
\begin{equation}
\label{eq:K}
K_F = \limsup_{k\to +\infty} |v_k| 
\end{equation}
with $\{v_k\}$ given by the iteration \eqref{exp-x}.

In particularly, when $d_k = d$ is  {a} constant, we have a periodic force $F(t)$, and 
\begin{equation}
\label{cons-solve1}
\left\{
\begin{aligned}
v_{2k+1}&=1 + (v_{2k}-1)e^{-d},\\
v_{2k+2}&=-1 + (v_{2k+1}+1)e^{-d}.
\end{aligned}
\right.
\end{equation}

\begin{prop}
Consider the equation \eqref{eq:1} with $F(t)$ defined by  \eqref{nomdefineF}, when $d_k = d > 0$ is a constant, we have
\begin{equation}
\label{eq:K0}
K_F = \limsup_{t\to\infty} |v(t; F)| = \dfrac{1 - e^{-d}}{1 + e^{-d}}.
\end{equation}
\end{prop}
\noindent\textbf{Proof} From \eqref{cons-solve1}, it is easy to obtain
$$
v_{2k+2} =  e^{-2d} v_{2k} - (1 - e^{-d})^2,\quad v_0 = 0,
$$
which gives
$$
v_{2k} = (e^{-2 k d } - 1) \dfrac{(1-e^{-d})}{(1 + e^{-d})},
$$
and, from  \eqref{cons-solve1},
$$
v_{2k+1} = 1 + e^{-d}\left((e^{-2 k d } - 1) \dfrac{(1-e^{-d})}{(1 + e^{-d})} - 1\right).
$$
Thus, when $k\to +\infty$, we have
$$
\lim_{k\to\infty}v_{2k} = -\dfrac{1 - e^{-d}}{1 + e^{-d}}, \quad\lim_{k\to\infty} v_{2k+1} = \frac{1 - e^{-d}}{1 + e^{-d}}.
$$
Hence, 
$$K_F = \limsup_{t\to\infty} |v(t; F)| = \max\{\lim_{k\to\infty}|v_{2k}|, \lim_{k\to\infty} |v_{2k+1}| \} =  \frac{1 - e^{-d}}{1 + e^{-d}}.$$

\begin{figure}[htbp]
\centering
\includegraphics[width=8cm]{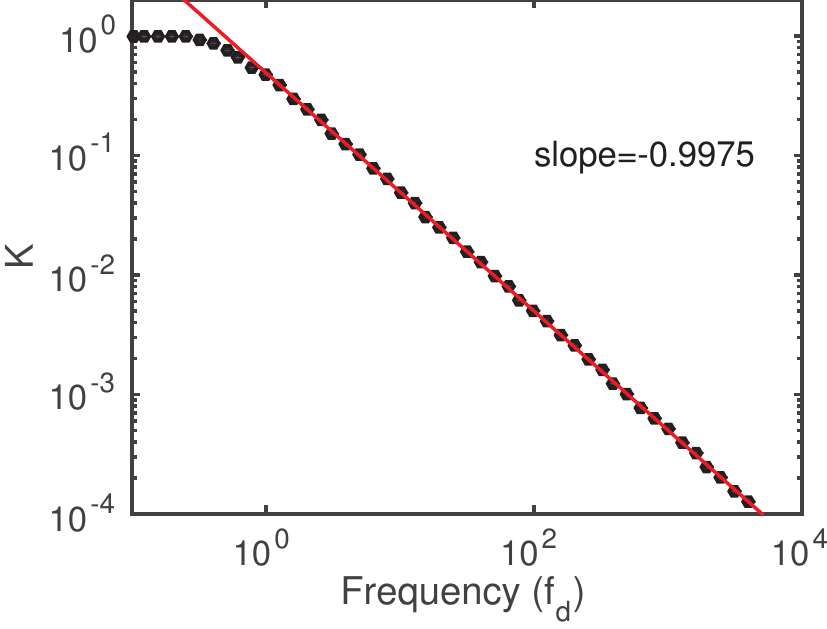}
\caption{Log-log plot for relationship between the upper bound $K_F$ and the frequency $f_d = 1/d$ when $d$ is a constant.}
\label{Fig-dconstent}
\end{figure}

The dependence function \eqref{eq:K0} is shown in Fig \ref{Fig-dconstent}. It is easy to see that, when the frequency $f_d = 1/d$ is small (or $d$ is large), the upper bound $K_F$ approaches to $1$. When the frequency $f_d$ is large ($f_d > 1$),  Fig \ref{Fig-dconstent} shows a power law dependence $K_F \sim f_d^{-1}$.  {Here, $f_d$ measures the frequency of changing the sign in the `random force' $F(t)$.}

\subsection{Deterministic Brownian motion generated from delay differential equation}

\begin{figure}[htbp]
\centering
\includegraphics[width=12cm]{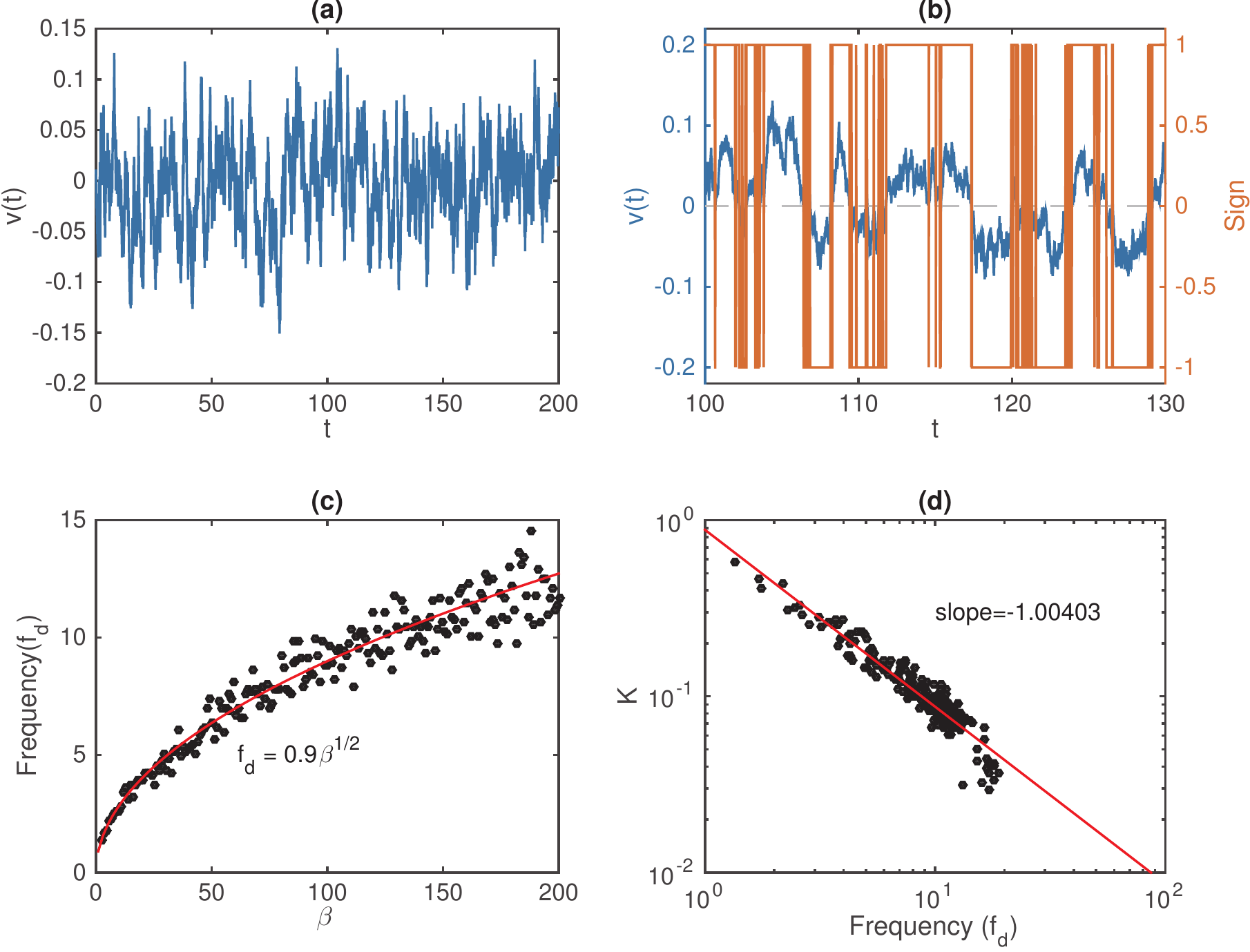}
\caption{Power law dependence of the delay differential equation model. (a) Solution of the equation \eqref{eq:dde1} with $\beta = 52$.  (b) Segment of the solution $v(t)$ in (a) and the corresponding sign switches. (c) Relation between the frequency $f_d$ of sign changes and the parameter $\beta$, red line shows the fitting of $f_d = 0.9 \beta^{1/2}$. (d) Log-log plot for relationship between the upper bound $K$ and the frequency.}
\label{Fig-dde}
\end{figure}

Now, we re-examine the delay differential equation \eqref{eq:dde1}.  Here, the force term $f(t) = \sin (2 \pi \beta v(t-1))$ is given by the chaotic solution $v(t)$, which switches between positive and negative.  In previous studies, we have seen that the upper bound $K$ changes with $\beta$ according to the power law $K \sim \beta^{-1/2}$ when  {$\beta \to \infty$}. To associate this power law dependence with the frequency of sign changes in the force $f(t)$, we vary the parameter $\beta$ over the range $\beta \in (3, 200)$ and solve the equation \eqref{eq:dde1} to obtain corresponding solutions $v(t; \beta)$ (Fig. \ref{Fig-dde}(a)). For each value $\beta$, the corresponding solution randomly changes between positive and negative over time, and hence $\mathrm{sign}(v(t; \beta))$ randomly switches between $+1$ and $-1$ (Fig. \ref{Fig-dde}(b)). We further find out the average frequency of changing the sign of $v(t)$, which is dependent on $\beta$ in a way $f_d \sim \beta^{1/2}$ (Fig. \ref{Fig-dde}(c)). Hence, we would have the power law dependence $K \sim f_d^{-1}$, which is verified by numerical simulations shown in Fig. \ref{Fig-dde}(d).

\subsection{Random sequences $\{d_k\}$}

Now, we consider the situations with random sequences $\{d_k\}$  {with} different distributions. From the iterative formula \eqref{exp-x}, $|v(t_k)|$ approaches $1$ when $d_k$ is large. Hence, it is trivial when $d_k$ can always take large values. We are more interested at the situations when $d_k$ is mostly small. To this end, we select a family of distributions so that the average  $\langle d\rangle$ may approach zero. 
  
\begin{table}[htbp]
\centering
\caption{Distributions of $\{d_k\}$}
{\begin{tabular}{|l|c|c|c|c|}
\hline
Distribution & Density function & Mean & Variance & CV\\
\hline
exponential distribution &$\lambda e^{-\lambda x}$ & $\frac{1}{\lambda}$ & $\frac{1}{\lambda^2}$ & $1$\\
\hline
gamma distribution & $\frac{\beta^\alpha}{\Gamma(\alpha)} x^{\alpha - 1} e^{-\beta x}$ & $\frac{\alpha}{\beta}$ & $\frac{\alpha}{\beta^2}$&  {$1/\sqrt{\alpha}$}\\
\hline
beta distribution &$\frac{\Gamma(\alpha+\beta)}{\Gamma(\alpha)\Gamma(\beta)}x^{\alpha - 1}(1-x)^{\beta -1}$ & $\frac{\alpha}{\alpha +\beta}$ & $\frac{\alpha \beta}{(\alpha+\beta)^2(\alpha + \beta + 1)}$ & $\sqrt{\frac{\beta}{\alpha (\alpha + \beta + 1)}}$\\
\hline
lognormal distribution &  $\frac{1}{\sqrt{2\pi} x \sigma} e^{-(\ln x - \mu)^2/(2 \sigma^2))}$& $e^{\mu + \sigma^2/2}$ & $e^{\sigma^2 - 1} e^{2 \mu + \sigma^2}$ & $\sqrt{e^{\sigma^2 - 1}}$ \\
\hline
\end{tabular}}
\end{table}

\begin{figure}[htbp]
      \centering
      \includegraphics[width=12cm]{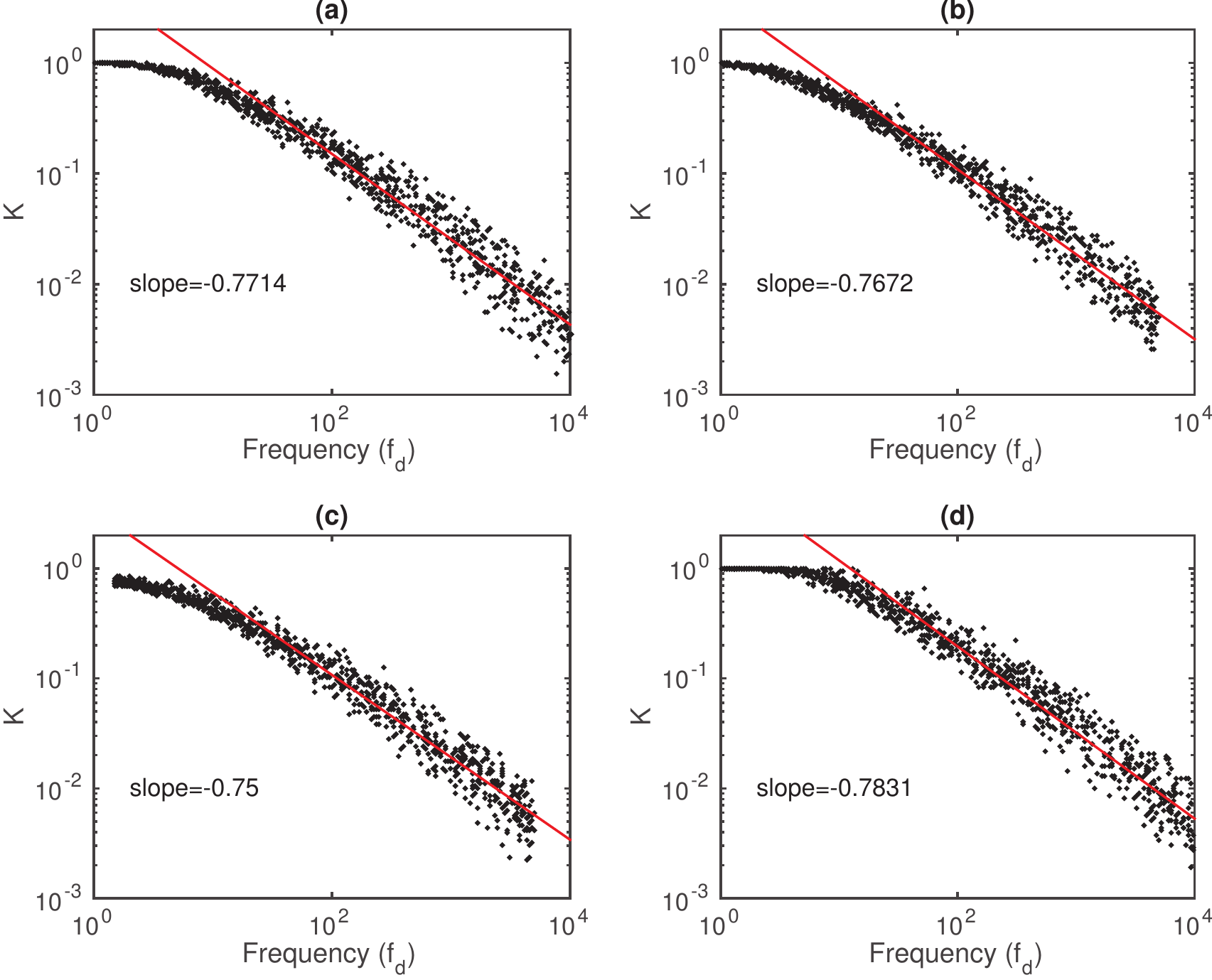}
      \caption{Power law dependence between the upper bound $K_F$ and the frequency $f_d$. Results show the power law dependences with various distributions for the random sequences $\{d_k\}$: (a) exponential distribution with $1<\lambda<1000$, (b) gamma distribution with $\alpha = 2$ and $1<\beta < 1000$; (c) beta distribution with $\alpha = 2$ and $1 < \beta < 1000$; (d) lognormal  {distribution} with $\sigma = 1$ and $-10 < \mu< -0.5$.}
      \label{Fig-expresult}
\end{figure}

Here, we consider four type distributions, which are shown in Table ~1.  According to the mean of these distributions, we can adjust the parameters to ensure that the average interval $\langle d \rangle$ approaches zero with various parameters.  For each distribution, we generate random sequences $\{d_k\}$ with given parameters, and find the upper bound $K_F$ according to \eqref{exp-x} and \eqref{eq:K}.  Fig. \ref{Fig-expresult} shows the log-log plots of the upper bound $K_F$ versus the frequency $f_d$ of the sequences $\{d_k\}$. It is obvious to see power law dependences for different distributions of the random sequences. 
 
\begin{figure}[htbp]
      \centering
      \includegraphics[width=12cm]{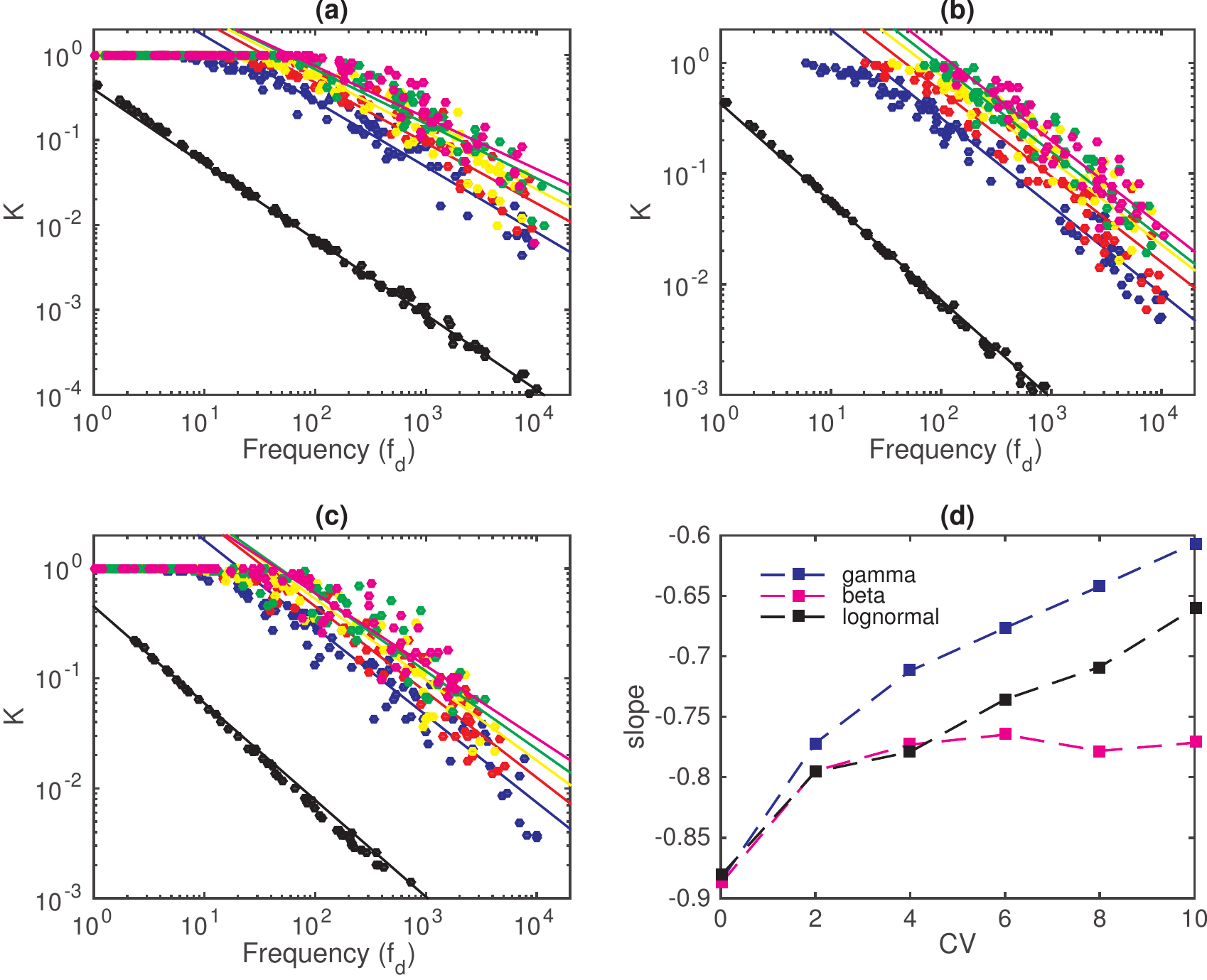}
      \caption{Power law dependences with different CV values. (a) Log-log plot for  {the} relationship between the upper bound $K_F$ and the frequency for gamma distribution $\{d_k\}$ with various CV. (b) Log-log plot for  {the} relationship between upper bound $K_F$ and the frequency for beta distribution $\{d_k\}$ with various CV. (c) Log-log plot for  {the} relationship between upper bound $K_F$ and the frequency for lognormal distribution $\{d_k\}$ with various CV. (d) Dependence of the  {slopes} in (a)-(c) and CV.  In (a)-(c), lines show the fitting with power law dependence, results obtained from different values CV  {($\mathrm{CV}=0.01, 2, 4, 6, 8, 10$)} are represented by different color.}
      \label{Fig-4}
\end{figure}

We note that the  {slopes} in Fig. \ref{Fig-expresult} are different from $-1$ as  {we have seen} in the above discussions of constant $d_k\equiv d$. This indicates that the  {slope} may dependences on the some other properties of the distribution. To check this dependence, we consider gamma distribution, beta distribution, and lognormal distribution sequences with varying  {the} coefficient of variation (CV). The coefficient of variation  {is often used to describe the fluctuation of a random process, and} is defined as  the ratio of  {the} standard deviation to  {the} mean, which are given in Table~1.  Results are shown in Fig. \ref{Fig-4}, in which power law dependence can be applied to fit the relationship between the upper bound $K_F$ and the frequency $f_d$ for random sequences $\{d_k\}$ with  {different} values CV. We note that the  {slopes} may depends on CV,  {which show increase with CV for the three type distributions}. Particularly, the slope approaches $-1$ when CV decreases  {to} $0$, which is consistent with the above result of constant sequences $d_k \equiv d$. 

\section{Frequency modulation and amplitude modulation}

Previous discussions shown that the upper bound of the equation 
\begin{equation}
\label{eq:18}
\dfrac{d v}{d t} = -\gamma v + F(t)
\end{equation}
can be modulated by the frequency of switching the sign of perturbation $F(t)$, in a way of power law dependence.  Thus, the equation of form \eqref{eq:18} can be served as a transformation from frequency modulation to amplitude modulation.  Particularly, for a dynamic system with bistable steady states separated by an energy barrier, we may control the frequency of steady state switches by regulating of the frequency in the perturbation function.   

\begin{figure}[htbp]
\centering
\includegraphics[width=12cm]{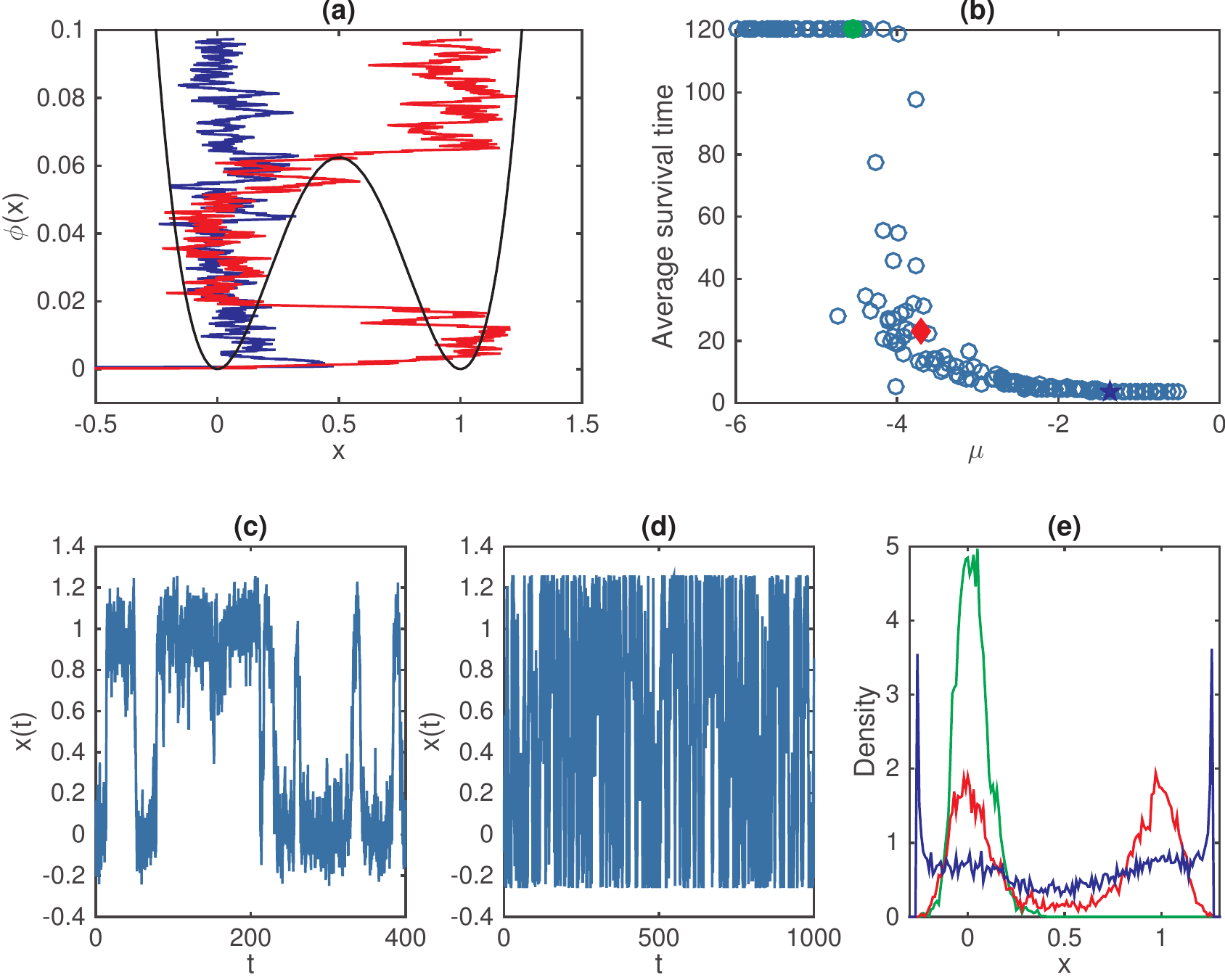}
\caption{Steady state switches induced by random perturbation. (a) Potential function. Blue and red lines show superposition of two sample solutions that either stay at one local minimal or switches between two local minimal states, respectively.  {Here, $\mu = -4.538$  for the blue line, and $\mu=-3.7401$ for the red line.} (b) Dependence of the average survival time of state $x^* = 0$ versus the parameter $\mu$ of the random sequence $\{d_k\}$. (c) Sample solution with $\mu$ taken from the red diamond in (b). (d) Sample solution with $\mu$ taken from the blue pentagram in (b). (e) Density functions of the solution trajectories $x(t)$ with different $\mu$ values from (b), green square, red diamond, and blue pentagram, respectively.}
\label{fig:5}
\end{figure}

To examine this idea, we consider the equation of form 
\begin{equation}
\label{eq:16}
\dfrac{d x}{d t} = - \phi'(x) + F(t),\quad \phi(x)  = x^2 - 2 x^3 + x^4,
\end{equation}
where the perturbation function $F(t)$ is defined as \eqref{nomdefineF}, with sequence $\{d_k\}$ taken from lognormal distribution with parameters $\sigma = 1$ and $-6.0 < \mu < -0.5$.  The potential function $\phi(x)$ has a form of double well, so that the equation \eqref{eq:16} has two steady states $x_1^* = 0$ and $x_2^* = 1$, which are separated by an energy barrier at $x = 0.5$ (Fig. \ref{fig:5}a).   

When there is a perturbation, while we linearize the equation \eqref{eq:16} at a steady state $x = x^*$, we obtained 
\begin{equation}
\label{eq:17}
\dfrac{d x}{d t} = - \gamma (x- x^*)  + F(t) + o(|x-x^*|).
\end{equation}
Previous discussions shown that the upper bound $|x(t) - x^*|$ of \eqref{eq:17} is determined by the frequency $f_d$ according to a power law dependence. Moreover, when the upper bound $|x(t) - x^*|$ is larger than the barrier $x = 0.5$, the state can switch from one state to the other one. Hence, we would expect to regulate the frequency of  {the switching behavior} between two steady states $x_1^*$ and $x_2^*$ by modulating the parameter $\mu$ in defining the random sequence $\{d_k\}$. 

 {From Table ~1, the mean value of $\{d_k\}$ is given by the parameter $\mu$ as $\langle d_k\rangle = e^{\mu + 1/2}$, hence, the frequency $f_d = 1/\langle d_k\rangle = e^{-(\mu +1/2)}$.}  We vary the parameter $-6 < \mu<-0.5$, so that the frequency $f_d$ changes over the range $1<f_d<e^{5.5}$. For each $\mu$, we generate a random sequence $\{d_k\}$ and solve the equation \eqref{eq:16} with initial condition $x(0 ) = 0$.  For each solution, we calculate the average survival time (AST) of the state $x_1^* = 0$, which are shown in Fig. \ref{fig:5}b (here $\mathrm{AST} = 120$ means no switches from $x_1^*$ to $x_2^*$). From Fig. \ref{fig:5}b, there is a threshold $\mu^*$, so that when $\mu < \mu^*$, switches from $x_1^*$ to $x_2^*$ do not occur during simulation, \textit{i.e.}, the state $x_1^*$ is stable under random perturbation. When $\mu > \mu^*$, the solution $x(t)$ can leave the state $x_1^*$, and the AST of $x_1^*$ decreases with the increasing of $\mu$ (decreasing of the frequency $f_d$).  Specifically, two sample solutions, with $\mu$ marked with red diamond and blue pentagram in (b), are shown in Fig. \ref{fig:5}c-d, respectively.  Moreover,  density of the solution trajectories $x(t)$ with different $\mu$ values are given in Fig. \ref{fig:5}e, which show the changes of the density function with random sequences $\{d_k\}$ in defining the perturbation function $F(t)$.   {These results suggest that the frequency of steady state switches can be regulated by modulating the parameter $\mu$ in defining the random sequence $\{d_k\}$.}

\section{Conclusion}

This paper studies the random equation of form \eqref{eq:1}, which is often applied to model a decay system with randomly switches forces.  We explore how the maximum displacement from the equilibrium state depends on the statistical properties of  {the time points of random force switches}. We show that the frequency of  {random forces switches} is essential for the maximum displacement, and there is a power law dependences between the upper bound and the frequency. The slope of  {the} power law dependence is dependent on different distributions of the random sequence $\{d_k\}$  {which is the base} to define the random perturbation.  Thus, the random equation of form \eqref{eq:1} can be served as a transformation from frequency modulation to amplitude modulation, and regulation of the perturbation frequency can be applied to  modulated the stability and state transition in systems with multiple steady states. 

Moreover, we re-examine the determined Brownian motion \eqref{eq:dde1} defined by nonlinear delay differential equations. We show that the frequency of sign changes  of the `chaotic' solution depends on the equation parameter $\beta$ in an order of $\beta^{1/2}$, and the upper bound of the solution depends on the frequency with a power  {law} $K \sim f_d^{-1}$, and hence we confirm the dependence $K \sim \beta^{-1/2}$ reported in previous studies\cite{LeiMackeyPRE2011}. 

\section*{Acknowledgments} 
This works was founded by National Natural Science Foundation of China (NSFC 11831015). JL thanks Professor Michael Mackey from McGill university for arising this interesting question and helpful discussions. 


\end{document}